\documentclass[12pt]{article}
\usepackage{amsfonts}
\usepackage{latexsym}
\usepackage{amsmath}
\usepackage{amssymb}
\usepackage{amssymb}
\usepackage{slashed}

\newcommand{\newsection}{    
\setcounter{equation}{0}\section}
\def\appendix#1{\addtocounter{section}{1}\setcounter{equation}{0}
\renewcommand{\thesection}{\Alph{section}}
\section*{Appendix \thesection\protect\indent \parbox[t]{11.15cm}{#1}}
\addcontentsline{toc}{section}{Appendix \thesection\ \ \ #1}}

\newcommand{\be}{\begin{eqnarray}}
\newcommand{\ee}{\end{eqnarray}}
\newcommand{\bea}{\begin{eqnarray}}
\newcommand{\eea}{\end{eqnarray}}
\newcommand{\ba}{\begin{array}}
\newcommand{\ea}{\end{array}}

\newenvironment{frcseries}{\fontfamily{frc}\selectfont}{}

\def\a{\alpha}
\def\b{\beta}

\def\d{\delta}

\def\D{\Delta}

\def\P{\Psi}

\newcommand{\p}{\phi}

\def\cA{{\cal A}}

\def\cL{{\cal L}}

\def\hp{{\bcc \phi}}

\def\hQ{{\hat{Q}}}

\def\hn{{\hat{\nu}}}

\def\bcc#1{\buildrel \circ \over #1}

\def\bcn {{\bcc \nabla}}

\def\rsq{{\cal{O}}(r^2)}

\def\cS{{\cal S}}
\def \tn {\tilde{\nabla}}
\def\w {\wedge}

\def\bcc#1{\buildrel \circ \over #1}

\def\hh{{\bcc h}}
\def\hhn{{\bcc \nabla}}
\def\hP{{\bcc \P}}
\def\hW{{\bcc W}}
\def\htH{{\bcc H}}
\def\hD{{\bcc \Delta}}

\def\td{\tilde{d}}
\def\hg{{\bcc g}}
\def\hQ{{\bcc Q}}

\def\hf{{\bcc f}}
\def\hX{{\bcc X}} 
\def\hg{{\bcc g}}

\def\tn {{\tilde{\nabla}}}

\def\dh {{\dot{h}}}

\def\hn {{\tilde{\nabla}}}

\begin{document}
\begin{titlepage}
\begin{center}
\vspace*{-1.0cm}
\hfill DMUS--MP--16/22 \\

\vspace{2.0cm} {\Large \bf Moduli Spaces of Transverse Deformations
of Near-Horizon Geometries} \\[.2cm]

\vskip 2cm
\textsc{A. Fontanella and J. B.  Gutowski} 
\\
\vskip .6cm

\begin{small}
\textit{Department of Mathematics,
University of Surrey \\
Guildford, GU2 7XH, UK. \\
Email: a.fontanella@surrey.ac.uk \\
Email: j.gutowski@surrey.ac.uk}
\end{small}\\*[.6cm]

\end{center}

\vskip 3.5 cm
\begin{abstract}

\vskip1cm

We investigate deformations of extremal near-horizon geometries in 
Einstein-Maxwell-Dilaton theory, including various topological terms,
and also in D=11 supergravity. By linearizing the field equations 
and Bianchi identities
over the compact spatial cross-sections of the near-horizon geometry, 
we prove that the moduli associated with such deformations are constrained 
by elliptic systems of PDEs. The moduli space of deformations
of near-horizon geometries in these theories is therefore shown to be 
finite dimensional.

\end{abstract}

\end{titlepage}


 \setcounter{section}{0}

\newsection{Introduction}

The relationship between near-horizon geometries and 
black hole solutions is of considerable interest. Although every
extremal black hole has a well-defined near-horizon geometry,
obtained by taking a certain decoupling limit of the black hole solution,
there is currently no way to definitively determine if a given near-horizon geometry
can be extended away from the near-horizon limit to produce a genuine black hole solution. 
There are also issues of uniqueness in higher dimensions.
The strong uniqueness theorems established in four dimensions \cite{israel, carter, hawking, robinson1, israel2, mazur, robinson} break down, and there exist different
black hole solutions with the same asymptotic charges, and also
different black hole solutions with the same near-horizon geometries.
For example,
in five dimensions, the near horizon geometry $AdS_3 \times S^2$ admits two possible extensions,
one to a supersymmetric black string \cite{Chamseddine:1999qs}, and the other to a supersymmetric black ring \cite{Elvang:2004rt}, though with different asymptotic conditions. 
There could also be other extensions in this case.

At the level of the near-horizon geometries, significant progress has been made in 
classifying such solutions, e.g. \cite{Kunduri:2006uh, Kunduri:2008rs, hethor, Grover:2013ima, Gutowski:2013kma}.
For non-supersymmetric solutions, the classifications assume the existence of sufficiently
many rotational $U(1)$ isometries on the spatial cross-sections of the event horizons.
For supersymmetric solutions, index theory techniques have been developed which constrain
the number of supersymmetries, and which in many cases imply that supersymmetry is enhanced.
The next step is to determine which near-horizon geometries actually correspond to a genuine
black hole, perhaps with a given asymptotic geometry. In principle,  this could be done by
using the fact that in the neighbourhood of a Killing horizon, the metric can be written 
in Gaussian Null co-ordinates \cite{isen, gnull}. There is a radial co-ordinate $r$,
such that the horizon is located at $r=0$, and the near-horizon solution corresponds
to taking the lowest order terms in $r$ in various components of the metric, which are assumed to be analytic in $r$. Determining the extension of the near-horizon solution amounts to solving the Einstein equations at higher and higher order in $r$. 
If the theory is also coupled to some matter, such as scalars or gauge fields, then these
also must be appropriately expanded out order by order in $r$, and the various field
equations solved at higher orders in $r$. In practice, this is extremely involved, 
and furthermore the Gausiann Null co-ordinate system in general will break down for sufficiently large $r$, so incorporating asymptotic data into the extension is difficult.
It is unknown if one can systematically classify obstructions to such an extension.

In the absence of a comprehensive understanding of the theory of extensions of
near-horizon geometries, we shall restrict ourselves to the first order problem.
By this we mean that we shall consider the first order terms in $r$ in the Taylor expansion
of the metric, and other matter fields, and view such terms as being small perturbations
of the near-horizon solution. The field equations are linearized with respect to these moduli, and we consider the issue of whether the space of moduli is finite dimensional. If the
moduli space were to fail to be finite dimensional, then this would imply very little
control over the possible extensions of near-horizon geometries. 

In this work, we shall investigate in particular the cases of Einstein-Maxwell-Dilaton theories
in any dimension,  including topological terms in the special cases $D=4,5$; and
also $D=11$ supergravity, for which the topological term coupling is kept arbitrary.
We show that the moduli space of first order radial deformations of any near-horizon solution of such theories is finite dimensional. The method we shall use is a development of
that which was first used to analyse the transverse moduli spaces of near-horizon solutions
in higher dimensional Einstein gravity coupled to a cosmological constant in \cite{Li:2015wsa},
and following that, in the case of minimal D=5 ungauged supergravity \cite{Dunajski:2016rtx}.
In the latter analysis, supersymmetry was used extensively in order to obtain
conditions on the gauge field strength. Here, we shall not make any use of supersymmetry,
and we work purely in terms of bosonic field equations. 

The analysis proceeds in two steps. First, some of the metric and matter field moduli are
fixed in terms of the other metric and matter field moduli, by making use of some of
the Bianchi identities and field equations. Secondly, once these moduli are fixed, elliptic systems of PDEs are obtained which constrain the remaining metric and matter moduli, again by exploiting some of the Bianchi identities and field equations. Assuming that the
spatial cross-sections of the near-horizon geometries are compact, this implies that the
moduli space associated with the moduli appearing in these elliptic equations is finite dimensional, via standard Fredholm theory. On making use of the moduli fixing conditions, this in turn implies that the
entire moduli space is finite dimensional. 

The plan of this paper is as follows. In section 2, we introduce our notation, and consider
in particular the metric transverse moduli. We also consider the gauge transformations
which act on these moduli, and recall the proof that a gauge can be chosen
for which the metric trace modulus lies in the kernel of an elliptic operator, which
decouples entirely from all the other moduli \cite{Li:2015wsa}. In sections 3 and 4 we
analyse the Einstein-Maxwell-Dilaton and D=11 supergravity theories respectively. For both theories, the
matter field moduli are found, the moduli fixing conditions for metric and matter moduli are obtained, and the elliptic
systems of PDEs on the remaining moduli are determined explicitly. In section 5 we present our conclusions.

\newsection{Metric Moduli}
 
In this section we briefly describe the metric moduli associated with radial deformations
of extremal near-horizon geometries. These moduli were first considered in the analysis of transverse deformations of extremal horizons in pure gravity with a cosmological constant 
constructed in \cite{Li:2015wsa}, and we summarize them here for convenience.

The metric moduli are common to all of the theories which we shall consider in the following sections. We assume that the black hole event horizon is a Killing
horizon, and use Gaussian Null co-ordinates $(u,r, y^A)$ adapted to the Killing horizon \cite{isen, gnull}. The metric
written in these co-ordinates is
\bea
ds^2 = 2du\Big(dr+rh-\frac{1}{2}r^2\Delta du\Big)+ ds^2_{\cal{S}}
\eea
where ${\partial \over \partial u}$ is an isometry of the geometry, which is null
on the Killing horizon at $r=0$. The metric
\bea
ds^2_{\cal{S}}=g_{AB} dy^A dy^B 
\eea
is the metric on spatial cross-sections of the geometry, which is $u$-independent and analytic in the radial co-ordinate $r$, and $\Delta$, $h$ are a scalar
and 1-form on ${\cal{S}}$. $\Delta$ and
$h$ are also $u$-independent, and analytic in $r$. 
We shall refer to $g$, $\Delta$ and $h$ as the horizon (metric) data. In what follows,
we shall take ${\bf{e}}^i$ to be an orthonormal basis for ${\cal{S}}$.
In what follows we shall find it particularly useful to choose a light-cone basis
\be
\label{nhbasis}
\mathbf{e}^+ = du \ , \qquad
\mathbf{e}^- = dr + rh_A dy^A - \frac{1}{2} r^2 \D du \ , 
\ee
with respect to which the spacetime metric is 
\be
\label{nearhormetr}
ds^2 = 2 \mathbf{e}^+ \mathbf{e}^- + ds^2_{\cal{S}} \ . 
\ee

To obtain the moduli associated with the horizon data, we simply Taylor expand the 
horizon data in $r$,
\bea
\Delta &=& \bcc{\Delta}(y)+ r \delta \Delta (y) + \rsq,
\nonumber \\
h &=& \bcc{h} (y) + r \delta h (y) + \rsq,
\nonumber \\
g &=& \bcc{g}(y)+ r \delta g(y) + \rsq
\eea
where $\bcc{\Delta}, \bcc{h}, \bcc{g}$ are the zeroth order terms, 
and the metric moduli are $\delta \Delta, \delta h, \delta g$. We shall
assume that ${\bcc{{\cal{S}}}}$, equipped with metric ${\bcc{g}}$, is compact.

The metric moduli admit a gauge transformation, associated with
infinitessimal diffeomorphisms, assumed to be of the same order as the metric moduli, generated by the vector field \cite{Li:2015wsa}
\bea
\xi = {1 \over 2} f \bigg(dr+r \bcc{h}-{1 \over 2} r^2 \bcc{\Delta} du \bigg)
-{1 \over 4} r^2 \bigg(\bcc{\Delta} f + {\cal{L}}_{\bcc{h}} f \bigg) du -{1 \over 2} r df
\eea
for an arbitrary smooth function $f$ on ${\cal{S}}$; with respect to which the 
horizon data transform as

\bea
\label{gtrans}
\delta {{g}}_{ij} &\rightarrow& \delta g_{ij} + \bcc{\nabla}_i \bcc{\nabla}_j f
- \bcc{h}_{(i} \bcc{\nabla}_{j)} f
\nonumber \\
\delta {{h}}_i &\rightarrow& \delta h_i +{1 \over 2} \bcc{\Delta} {\bcc{\nabla}}_i f
-{1 \over 4} ({\bcc{\nabla}}_i \bcc{h}_j) {\bcc{\nabla}}^j f
-{1 \over 4} \bcc{h}_i \bcc{h}_j {\bcc{\nabla}}^j f
+{1 \over 2} ({\bcc{\nabla}}_j \bcc{h}_i) {\bcc{\nabla}}^j f
+{1 \over 4} \bcc{h}_j {\bcc{\nabla}}_i {\bcc{\nabla}}^j f
\nonumber \\
\delta {{\Delta}} &\rightarrow& \delta \Delta +{1 \over 2} {\bcc{\nabla}}^i f
\bigg({\bcc{\nabla}}_i \bcc{\Delta} - \bcc{h}_i {\bcc{\Delta}} \bigg) \ ,
\eea
where indices $i,j,\dots$ are with respect to the orthonormal basis ${\bf{e}}^i\big|_{r=0}$
on ${\bcc{{\cal{S}}}}$, and $\bcn$ denotes the Levi-Civita connection on ${\bcc{{\cal{S}}}}$.

Before considering the moduli space calculations of the Einstein-Maxwell-Dilaton and
D=11 supergravity theories, we shall first consider the trace modulus $\delta g_k{}^k$.
This is constrained by an elliptic
PDE which decouples from the matter content, and is common to all of the theories. To see
this note that under the transformation ({\ref{gtrans}})
\bea
\label{trtrans}
\delta g_k{}^k \rightarrow \delta g_k{}^k + {\cal{D}} f
\eea
where ${\cal{D}}$, and its adjoint ${\cal{D}}^\dagger$, are given by
\bea
{\cal{D}} \equiv \bcn^2 - {\bcc h}^i \bcn_i \ , \qquad {\cal{D}}^\dagger = \bcn^2 + {\bcc h}^i \bcn_i + \bcn^i {\bcc h}_i \ .
\eea
We decompose $\delta g_k{}^k$ as a sum of two terms, $\phi \in {\rm Im} {\cal{D}}$, and $\phi^\perp \in \big({\rm Im} {\cal{D}} \big)^\perp$ as
\bea
\delta g_k{}^k = \phi + \phi^\perp \ .
\eea
Therefore
\bea
\phi = {\cal{D}} (\tau), \qquad {\cal{D}}^\dagger \phi^\perp =0
\eea
where $\tau$ is a smooth function. On setting $f=-\tau$ in ({\ref{trtrans}}), we have
\bea
\delta g_k{}^k = \phi^\perp
\eea
and hence
\bea
\label{trelliptic}
\bigg(\bcn^2 + {\bcc h}^i \bcn_i + \bcn^i {\bcc h}_i \bigg) \delta g_k{}^k=0 \ .
\eea
This is an elliptic PDE. This condition is independent of the matter content of the
theory which we couple to gravity. We shall make use of this result in the analysis of the
metric moduli in the following sections, in which we consider various different theories, and their associated moduli spaces. In particular, the linearized Einstein equations include
a Hessian term in $\delta g_k{}^k$. Without the gauge fixing condition, this term
would destroy the ellipticity of the associated equation. However, as the trace modulus
is fixed by ({\ref{trelliptic}}), the linearized Einstein equation acting on the traceless part of $\delta g$ will be elliptic.

\newsection{Horizons in Einstein-Maxwell-Dilaton theory}

In this section, we consider the moduli space associated with the Einstein-Maxwell-Dilaton
theory in $D$ dimensions. The theory has a $D$-dimensional metric ${\hat{g}}$, 
as well as Abelian gauge 2-form field strengths $F^I=dA^I$, $I=1, ... , N$ and uncharged scalars $\phi^a$, $a=1,..., M$, governed by the action: 
\be
\label{EMDaction}
S = \int d^D x \sqrt{-g} \left( R - \frac{1}{2} f_{ab}(\phi) \nabla_{\mu}\phi^a \nabla^{\mu} \phi^b - V(\phi) - \frac{1}{4} Q_{IJ}(\phi) F_{\mu\nu}^I F^{J\, \mu\nu} \right) \ , 
\ee
where the couplings $f_{ab}$ and $Q_{IJ}$ are functions of the scalar fields $\phi^a$. 

The Einstein, gauge and scalar field equations are: 
\be
\label{einst}
R_{\mu\nu} 
- \frac{1}{2} f_{ab} \nabla_{\mu}\phi^a \nabla_{\nu}\phi^b
- \frac{1}{2} Q_{IJ}F^I_{\mu\alpha}F^{J\,}_{\nu}{}^{\alpha}
+ \frac{1}{D-2} {\hat{g}}_{\mu\nu}\bigg( \frac{1}{4}Q_{IJ} F_{\alpha\beta}^I F^{J\, \a\b} - V \bigg) = 0  \ ,
\ee
and
\be
\label{gauge}
\nabla^{\nu} \left(Q_{IJ} F^J_{\nu\mu} \right) = 0 \ ,
\ee
and 
\be
\label{scalar}
f_{ab} \nabla_{\mu}\nabla^{\mu} \phi^b 
+ (\partial_b f_{ac} - \frac{1}{2} \partial_a f_{bc}) \nabla_{\mu}\phi^b \nabla^{\mu} \phi^c 
- \partial_a V  - \frac{1}{4} \partial_a Q_{IJ} F^I_{\mu\nu} F^{J\, \mu\nu} = 0 \ ,
\ee
where $\partial_a \equiv \partial / \partial \phi^a$. The Bianchi identity is
\bea
dF^I=0 \ .
\eea
We furthermore assume that ${\partial \over \partial u}$ is a symmetry of the full solution, i.e.
\be
\cL_{\partial_u} A^I = 0 \ , \qquad\qquad \cL_{\partial_u} \phi^a = 0 \ . 
\ee
The scalar fields $\phi^a$  depend only on $r$ and $y^A$, and the field strengths $F^I$ are written in the Gaussian Null co-ordinates as
\be 
F^I= \P^I du\w dr + r  W^I_A du \w dy^A + Z^I_A dr \w  dy^A + {1 \over 2} \tilde{F}^I_{AB} dy^A\w dy^B \ , 
\ee
where $\P^I$ is a $u$-independent scalar, $W^I, Z^I$ are $u$-independent 1-forms, and $\tilde{F}^I$ is a $u$-independent 2-form defined on $\cS$. In particular, when written in the light-cone
basis, the field strengths are
\be
F^I = \P^I \mathbf{e}^+ \w \mathbf{e}^- + r \mathbf{e}^+ \w \left(W^I - \P^I h + \frac{1}{2} r \D Z^I \right)  + \mathbf{e}^- \w Z^I + H^I\ ,
\ee
where 
\be
H^I = \tilde{F}^I - r h \w Z^I  \ .
\ee
The Bianchi identity imposes the following set of conditions on the $F^I$ components
\be
\notag
&& \tilde{d}W^I = 0 \ , \qquad \tilde{d} \P^I - W^I - r \dot{W}^I = 0 \ , \\
\notag
&&\tilde{d} Z^I - \dot{H}^I - h\w Z^I - r \dot{h}\w Z^I - r h\w \dot{Z}^I = 0 \ , \\
&& \tilde{d} H^I + r \tilde{d}h \w Z^I - r h \w \tilde{d}Z^I = 0   \ , 
\ee
where we denote by $\tilde{d}$ the exterior derivative restricted to
hypersurfaces $r=const$, and by $\dot{\xi}$ the Lie derivative of $\xi$ along the vector field ${\partial \over \partial r}$, i.e. 
\be
\dot{\xi} \equiv \cL_{{\partial \over \partial r}} \xi  \ .
\ee
The decomposition of (\ref{einst}), (\ref{gauge}) and (\ref{scalar}) in terms of the horizon data is given in Appendix \ref{reducedEMD}. 

\subsection{Moduli space computation}
\label{moduli computation}
To begin the moduli space computation, we
consider the moduli associated with the scalar fields and the components
of the gauge field strengths.

The scalar fields are Taylor expanded as
\bea
\phi^a = {\bcc{{\phi^a}}} + r \delta \phi^a + {\cal{O}}(r^2)
\eea
The $\Psi^I$, $W^I$ and $H^I$ components of $F^I$ are expanded as
\bea
\Psi^I &=& {\bcc{{\Psi^I}}} + r \delta \Psi^I + {\cal{O}}(r^2)
\nonumber \\
W^I &=&  {\bcc{{W^I}}} + r \delta W^I + {\cal{O}}(r^2)
\nonumber \\
H^I &=& {\bcc{{H^I}}}+ r \delta H^I + {\cal{O}}(r^2)
\eea

There is however a subtlety with respect to the $Z^I$ components.
The $Z^I$ terms appear in the $F^I$ as $dr \wedge Z^I$,
which scales linearly with $r$, and hence we expand $Z^I$ as
\bea
Z^I = \delta Z^I + {\cal{O}} (r)
\eea
So, with these expansions, the scalar moduli are $\delta \phi^a$, and the gauge field moduli are
$\{\delta \Psi^I, \delta W^I, \delta Z^I, \delta H^I \}$.

The above expansions are chosen to be consistent with taking the near-horizon limit. To take the near-horizon limit, we take 
\be
\label{diffeo}
(u, r, y^A) \longrightarrow (\epsilon^{-1}u, \epsilon r , y^A ) \ , \qquad\qquad
\epsilon \in \mathbb{R}_{>0} \ .
\ee
On making this transformation, we note that all moduli terms (metric, scalar and gauge)
are linear in $\epsilon$. The near-horizon limit is obtained by taking $\epsilon \rightarrow 0$.
Our choice of $Z^I$ moduli expansion is also required for consistency with the $--$ component of the
Einstein equation.

Having determined the moduli expansions, we start to analyse the conditions imposed on
the moduli. From the Bianchi identity, we find the following conditions:
\be
\label{Bianchi_lin1}
\tilde{d}\d \P^I = 2 \d W^I \ ,  \\
\label{Bianchi_lin2}
 \tilde{d}\d Z^I -   \hh \w \d Z^I - \d H^I = 0 \ . 
\ee

At this stage we shall specify which of the moduli are fixed by Bianchi identities and
field equations.
Equation (\ref{Bianchi_lin1}) fixes $\d W^I$ in terms of $\d\P^I$, and (\ref{Bianchi_lin2}) fixes $\d H^I$ in terms of $\d Z^I$. 
By using the $-$ component of the gauge field equation (\ref{gauge_2}), we fix $\d \Psi^I$ as follows
\be
\label{psisubs}
\d \P^I = \hhn^i \d Z^I_i - \hh^i \d Z^I_i + \hQ^{IK} \partial_a \hQ_{KJ} \hhn^i \hp^a \d Z^J_i - \frac{1}{2} \hP^I \d g_k{}^k - \hQ^{IK} \partial_a \hQ_{KJ} \hP^J \d \p^a  \ , 
\ee
By using the $-i$ and the $+-$ components of the Einstein equations, (\ref{EMDeins-i}) and (\ref{EMDeins+-}), we fix $\d h$ and $\d \Delta$ respectively as follows
\be
\label{hsubs}
\notag
\d h_i &=&\frac{1}{2} \hhn_i \d g_k{}^k - \frac{1}{2} \hhn^j \d g_{ji}  - \frac{1}{4} \hh_i \d g_k{}^k + \frac{1}{2} \d g_{ij} \hh^j  \\ 
&+& \frac{1}{2} \hf_{ab} \d \phi^a \hhn_i \hp^b 
+ \frac{1}{2} \hQ_{IJ} ( \hP^I \d Z^J_i + \d Z^{I\, j} \htH^J_{ij} ) \ , 
\ee
and
\be
\label{deltsubs}
\notag
\d \D &=& \frac{1}{3} \hhn_i \d h^i + \frac{1}{12} \hh^i \hhn_i \d g_k{}^k - \hh^i \d h_i - \frac{1}{6} \hD \d g_k{}^k - \frac{1}{12} \hh_i \hh^i \d g_k{}^k + \frac{1}{3} \d g_{ij} \hh^i\hh^j \\
\notag
&-& \frac{1}{6} \d g_{ij} \hhn^i \hh^j - \frac{1}{6} \hh^j \hhn^i \d g_{ij} +\frac{1}{6(D-2)}\d \p^a \partial_a {\bcc Q_{IJ}}  (2 \htH^I_{ij} \htH^{J\, ij} + (D-3) \hP^I \hP^J ) \\
\notag
&+& \frac{1}{3(D-2)} \hQ_{IJ} ( 2 \d H^I_{ij} \htH^{J\, ij} - 2 \htH^{I\, \ell_1}{}_{j} \htH^{J\, \ell_2 j} \d g_{\ell_1\ell_2} + (D-3) \hP^I \d \P^J ) \\ 
&-& \frac{1}{3(D-2)} \d \p^a \partial_a {\bcc V} + \frac{4-D}{6(D-2)} \hQ_{IJ} (\hW^I_i \d Z^{J\, i} - \hP^I \hh^i \d Z^J_i ) \ .
\ee
So, the $\delta W^I, \delta H^I, \delta \Psi^I, \delta h, \delta \Delta$ moduli are fixed
in terms of the moduli $\delta Z^I, \delta g, \delta \phi^a$. We remark
that the moduli $\delta \Psi^I, \delta h, \delta H^I$ are linear in $\delta Z^I, \delta g, \delta \phi^a, \bcn \delta Z^I, \bcn \delta g, \bcn \delta \phi^a$, whereas $\delta W^I, \delta \Delta$
involve some second order derivative terms on  $\delta Z^I, \delta g, \delta \phi^a$.

We now turn to the remaining moduli $\delta Z^I, \delta g, \delta \phi^a$. We shall use
the Bianchi identities and field equations to construct elliptic systems of PDEs 
constraining these moduli, making use of the moduli fixing 
conditions on $\delta W^I, \delta \Psi^I, \delta h, \delta \Delta$. 

We start with the $\delta Z^I$ moduli. On taking the divergence of (\ref{Bianchi_lin2}), we obtain 
\be
\label{divBianchi}
\notag
\hhn^j\hhn_j \d Z^I_i &-& \hhn_i \hhn^j \d Z^I_j - {\bcc R}_{ij} \d Z^{I\, j} - \d Z^I_i \hhn_j \hh^j \\
&-& \hh^j \hhn_j \d Z^I_i + \d Z^{I\, j} \hhn_j \hh_i + \hh_i \hhn^j \d Z^I_j - \hhn^j \d H^I_{ji} = 0 \ ,
\ee
where $\bcc{R}$ denotes the Ricci tensor of ${\bcc{{\cal{S}}}}$.  
Using the $i$ component of the gauge field equation (\ref{gauge_3}), we express $\hhn^j \d H^I_{ji}$ as
\be
\notag
\hhn^j\d H^I_{ji} &=& - \hhn_i \hhn^j \d Z^I_j + \hhn_i ( \hh^j \d Z^I_j ) - \hQ^{IK} \partial_a \hQ_{KJ} \hhn_i (\hhn^j \hp^a \d Z^I_j ) \\
\notag
&+& \frac{1}{2} \hhn_i ( \hP^I \d g_k{}^k ) + \hQ^{IK} \partial_a \hQ_{KJ}
 \hhn_i ( \hP^J \d \p^a ) + \d g_{\ell_1\ell_2} \hhn^{\ell_1} \htH^{I\, \ell_2}{}_i + \hhn^j \d g_{jk} \htH^{I\, k}{}_i  \\
 \notag
&-&  \frac{1}{2} \hhn_j \d g_k{}^k \htH^{I\, j}{}_i  + \frac{1}{2} \hhn_j \d g^k{}_i \htH^{I\, j}{}_k + \frac{1}{2} \hhn_i \d g_j{}^k \htH^{I\, j}{}_k - \frac{1}{2} \hhn_k \d g_{ij} \htH^{I\, jk}  \\
\notag
&+& 2 \hh_i \d \P^I + 2 \hP^I \d h_i + 2\hh^j \d H^I_{ji} + 2\d h^j \htH^I_{ji} - \d\p^b \partial_b {\bcc Q^{IK}} \partial_a \hQ_{KJ} \hhn^j \hp^a \htH^J_{ji} \\
\notag
&-& \d \p^b\hQ^{IK}\partial_a \partial_b{\bcc Q_{KJ}} \hhn^j \hp^a \htH^J_{ji} - \hQ^{IK} \partial_a \hQ_{KJ} ( \htH^J_{ji}\hhn^j \d \p^a  + \d H^J_{ji} \hhn^j \hp^a )  \\
\notag
&+& \hQ^{IK} \partial_a \hQ_{KJ} \d g_{\ell_1\ell_2} \hhn^{\ell_1} \hp^a \htH^{J\, \ell_2}{}_i  + \hQ^{IK} \partial_a \hQ_{KJ} \d \p^a ( \hP^J \hh_i - \hW^J_i + \hh^j \htH^J_{ji} )   \\
\notag
&-& 2 \hD \d Z^I_i - \frac{1}{2} \d g_k{}^k ( \hW^I_i - \hP^I \hh_i ) + \d g_{ij} ( \hW^{I\, j} - \hP^I \hh^j ) \\
&+&\frac{1}{2} \d g_k{}^k \hh^j \htH^I_{ji} - 2 \hh^j \htH^{I\, k}{}_i \d g_{jk} - \hh_j \htH^{I\, jk} \d g_{ki} \ . 
\ee
Then we substitute this expression into ({\ref{divBianchi}}); the
$\hhn_j \hhn^i \d Z^I_i$ terms cancel out. Furthermore, the terms linear in
$\delta H^I$, $\delta h$ and $\delta \Psi^I$ are rewritten using 
the Bianchi identity ({\ref{Bianchi_lin2}}) and
({\ref{hsubs}}), ({\ref{psisubs}}), producing
terms linear in $\delta Z^I, \delta g, \delta \phi^a, \bcn \delta Z^I, \bcn \delta g, \bcn \delta \phi^a$. The resulting expression produces
an elliptic PDE for $\delta Z^I$, with principal symbol generated by $\bcn^2$.

Next, we consider the scalar moduli $\delta \phi^a$. The linearized scalar field equation is
\be 
\label{lin_sc}
\notag
&&\hhn_i \hhn^i \d \p^a - \d g_{ij} \hhn^i\hhn^j \hp^a - \hhn^i \d g_{ij} \hhn^j \hp^a + \frac{1}{2} \hhn^i \hp^a \hhn_i \d g_k{}^k - \d h^i \hhn_i \hp^a - \hh^i \hhn_i \d \p^a\\
\notag
 &&+ \d g_{ij} \hh^i \hhn^j \hp^a + \d\p^e\left[ \partial_e{\bcc f}^{ab}( \partial_c \hf_{bd} - \frac{1}{2}\partial_b\hat{f}_{cd}) + \hf^{ab}( \partial_c \partial_e\hf_{bd} - \frac{1}{2} \partial_b \partial_e\hf_{cd}) \right] \hhn_i \hp^c \hhn^i \hp^d   \\
\notag
&& - \hf^{ab} (\partial_c \hf_{bd} - \frac{1}{2} \partial_b \hf_{cd} ) \d g_{ij} \hhn^i \hp^c \hhn^j \hp^d + (2 \hf^{ab} \partial_{(c} \hf_{d) b}- \hf^{ab} \partial_b \hf_{cd}) \hhn_i \hp^c \hhn^i \d \p^d   \\
\notag
&&+ \frac{1}{4}\d \p^c ( \partial_c \hf^{ab} \partial_b \hQ_{IJ}  + \hf^{ab} \partial_b \partial_c \hQ_{IJ} ) ( 2 \hP^I \hP^J - \htH^I_{ij} \htH^{J\, ij} )  \\
\notag
&&+ \frac{1}{2} \hf^{ab} \partial_b \hQ_{IJ} ( 2 \hP^I \d \P^J - \htH^I_{ij} \d H^{J\, ij} + \d g_{\ell_1\ell_2} \htH^{I\, \ell_1}{}_k \htH^{J\, \ell_2 k} )  \\
\notag
&&- \d\p^c (\partial_c \hf^{ab} \partial_b {\bcc V} + \hf^{ab} \partial_b \partial_c{\bcc V}) + \hhn^j \hp^a ( \hh^i \d g_{ij} - \frac{1}{2}\d g_k{}^k \hh_j )  \\
\notag
&& + \d \p^a ( 2 \hD + \hh_i \hh^i - \hhn^i \hh_i )  - \d \p^c\hf^{ab}(\partial_c \hf_{bd} + \partial_d \hf_{bc} - \partial_b \hf_{cd} )   \hh^i \hhn_i \hp^d \\
&& - 2 \hh_i \hhn^i \d \p^a - \frac{1}{2} \d Z^I_i \hf^{ab} \partial_b \hQ_{IJ} (  \hW^{J\, i} - \hP^J \hh^i ) = 0  \ .
\ee
The terms linear in
$\delta H^I$, $\delta h$ and $\delta \Psi^I$ are again rewritten using 
the Bianchi identity ({\ref{Bianchi_lin2}}) and
({\ref{hsubs}}), ({\ref{psisubs}}), producing
terms linear in $\delta Z^I, \delta g, \delta \phi^a, \bcn \delta Z^I, \bcn \delta g, \bcn \delta \phi^a$. The resulting expression produces
an elliptic PDE for $\delta \phi^a$, with principal symbol generated by $\bcn^2$.

Lastly, we consider the metric moduli $\delta g$. The linearized $ij$ component of the Einstein equations is
\be
\label{ij_lin}
\hhn^2 \d g_{ij} - \hhn_i \hhn_j \d g_k{}^k - ( \hhn_{\ell} \hhn_j - \hhn_j \hhn_{\ell} ) \d g^{\ell}{}_i - ( \hhn_{\ell} \hhn_i - \hhn_i \hhn_{\ell} ) \d g^{\ell}{}_j = \cA_{ij} \ , 
\ee
where
\be
\notag
\cA_{ij} &=& ( \hhn_i \d g^k{}_j + \hhn_j \d g^k{}_i - \hhn^k \d g_{ij} ) \hh_k -  \hhn_{(i} ( \hh_{j)} \d g_k{}^k ) + 2\hhn_{(i} (\d g_{j)k} \hh^k)  \\
\notag
&+& 2\hhn_{(i} (\hhn_{j)} \hp^b \hf_{ab} \d \p^a )  + 2\hhn_{(i} (\d Z^J_{j)}  \hQ_{IJ} \hP^I ) + 2 \hhn_{(i} ( \htH^J_{j)k}\hQ_{IJ} \d Z^{I\, k}  ) \\
\notag
&-& 8\hh_{(i} \d h_{j)} + 2( - \hD + \frac{1}{2} \hhn_k \hh^k - \hh_k \hh^k ) \d g_{ij} \\
\notag
&-& 2 \d g_{(i}{}^k \hhn_{|k|} \hh_{j)} - 2 \hh^k \hhn_{(i} \d g_{j)k} + 2 \hh^k \hhn_k \d g_{ij} - 2 \hh_{(i} \hhn^k \d g_{j)k} \\
\notag
&+& 4 \hh_k \hh_{(i} \d g_{j)}{}^k + 2 \hh_{(i} \hhn_{j)} \d g_k{}^k + \d g_k{}^k ( \hhn_{(i}\hh_{j)} - \hh_i \hh_j ) - \d \p^c\partial_c \hf_{ab}  \hhn_i \hp^a \hhn_j \hp^b \\
\notag
&-& 2 \hf_{ab} \hhn_i \d \p^a \hhn_j \hp^b - \d \p^a\partial_a \hQ_{IJ}  \htH^I_{ik} \htH^J_j{}^k + \hQ_{IJ} \d g_{k\ell} \htH^I_i{}^k \htH^J_j{}^{\ell} + 2 \hQ_{IJ}  \d H^I_{k(i} \htH^J_{j)}{}^k \\
\notag
&-& \frac{2}{D-2} \hg_{ij} \d \p^a \partial_a {\bcc V} + \frac{1}{2(D-2)} \hg_{ij} \d \p^a \partial_a \hQ_{IJ} ( \htH^I_{k\ell} \htH^{J\, k\ell} - 2 \hP^I \hP^J ) \\
\notag
&+& \frac{1}{2(D-2)} \d g_{ij} \hQ_{IJ} ( \htH^I_{k\ell} \htH^{J\, k\ell} - 2 \hP^I \hP^J ) -{2 \over D-2} \d g_{ij} {\bcc V}
\\ \notag
&+& \frac{1}{D-2} \hg_{ij} \hQ_{IJ} ( \d H^I_{k\ell} \htH^{J\, k\ell} - \d g_{mn} \htH^{I\, m}{}_k \htH^{J\, nk} - 2 \d \P^I \hP^J ) +  2 \d \p^a \hf_{ab} \hh_{(i} \hhn_{j)} \hp^b   \\
&+& \frac{2}{D-2} \hg_{ij} \hQ_{IJ} \d Z^I_\ell ( \hW^{J\, \ell} - \hP^J \hh^\ell ) - 2 \hQ_{IJ} (\hW^I_{(i} \d Z^J_{j)} - \hP^I \d Z^J_{(j} \hh_{i)} ) \ ,
\ee
which is a linear expression in $\d Z^I, \d \p^a, \d g, \hhn \d Z^I, \hhn \d \p^a, \hhn \d g$. Furthermore, in ({\ref{ij_lin}}), terms of the form $( \hhn_{\ell} \hhn_j - \hhn_j \hhn_{\ell} ) \d g^{\ell}{}_i$ can be rewritten as terms linear in $\delta g$ and the Riemann tensor ${\bcc R}$, hence can be incorporated into the algebraic term on the RHS. The trace term $\delta g_k{}^k$
is fixed by the elliptic condition ({\ref{trelliptic}}), so ({\ref{ij_lin}})
is an elliptic set of PDEs for the traceless part of $\delta g$, with principal symbol generated by $\bcn^2$.

Taken together the conditions ({\ref{divBianchi}}), ({\ref{lin_sc}}), ({\ref{ij_lin}})
and ({\ref{trelliptic}}) constitute elliptic PDEs on the moduli $\delta Z^I$, $\delta \phi^a$, $\delta g$. The remaining moduli $\{\delta W^I, \delta H^I, \delta \Psi^I, \delta h, \delta \Delta \}$ are fixed in terms of $\{\delta Z^I, \delta \phi^a, \delta g \}$
by ({\ref{Bianchi_lin1}}), ({\ref{Bianchi_lin2}}), ({\ref{psisubs}}), ({\ref{hsubs}}) and ({\ref{deltsubs}}). The moduli space is therefore finite dimensional.

\subsection{Including Topological Terms}
In this section we shall add a topological term to the Einstein-Maxwell-Dilaton action (\ref{EMDaction}), and analyse explicitly in four and five dimensions any change to the analysis reported in section \ref{moduli computation}. Such cases are of particular relevance in the context of four and five-dimensional supergravity theories. We shall show in all cases that the addition of the topological terms does not affect the principal symbol of the systems of PDEs. 
Therefore we prove that the finiteness of the moduli space is also true in the presence of a topological term in the action.  

\subsubsection{Topological Terms in $D=4$} 

In $D=4$, we shall consider the following topological term
\be
S_{top} = \int t_{IJ}(\p) F^I \w F^J\ , 
\ee
where $t_{IJ}(\p)$ are functions of the scalar fields. 
The gauge field equation, in the presence of topological term, becomes
\be
\label{gauge_top4d}
\nabla^{\nu} ( Q_{IJ}(\p) F^J_{\nu\mu}) +   \epsilon_{\mu\nu\rho\sigma}F^{J\, \rho\sigma}\nabla^{\nu}t_{IJ}(\p ) = 0 \ .    
\ee
and the scalar field equation becomes
\be
\label{scalar_top4d}
\notag
f_{ab} \nabla_{\mu}\nabla^{\mu} \phi^b 
&+& (\partial_b f_{ac} - \frac{1}{2} \partial_a f_{bc}) \nabla_{\mu}\phi^b \nabla^{\mu} \phi^c 
- \partial_a V  \\
&-& \frac{1}{4} \partial_a Q_{IJ} F^I_{\mu\nu} F^{J\, \mu\nu} + \frac{1}{4}F^I_{\mu\nu}F^J_{\rho\sigma} \epsilon^{\mu\nu\rho\sigma} \partial_a t_{IJ} = 0 \ ,
\ee
The changes provided by the topological term relevant to the moduli space computation are here presented. The $-$ component of the gauge field equation, which enters into the
moduli fixing for $\delta \Psi^I$, is modified by:
\be
\epsilon^{+-ij} \bigg( \partial_a t_{IJ} \dot{\p}^a ( H^J_{ij} + 2rh_i Z^J_j) - 2 Z^J_j \partial_a t_{IJ} \hn_i \p^a \bigg)  \ , 
\ee
which is linear on the moduli $\d \p^a$ and $\d Z^I$. 
The $i$ component of the gauge field equation, which is used to construct the
elliptic system for $\delta Z^I$, is modified by:
\be
2 \epsilon^{+-ij} \bigg( \P^J \partial_a t_{IJ} \hn_j \p^a - r W^J_j \partial_a t_{IJ} \dot{\p}^a \bigg)  \ , 
\ee
which when linearized includes terms terms linear in $\delta Z^I$, $\delta \Psi^I$, $\delta \phi^a$. On eliminating the $\delta \Psi^I$ term, the extra terms
can be rewritten as terms linear in $\delta Z^I, \delta g, \delta \phi^a, \bcn \delta Z^I$, which do not affect the ellipticity of the resulting
system.

The scalar field equation is modified by:
\be
\epsilon^{+-ij} \partial_a t_{IJ} \bigg( \P^I H^J_{ij} + 2r Z^J_j (\P^I h_i  -W^I_i ) \bigg) \ , 
\ee
and includes terms linear in $\delta Z^I$, $\delta \Psi^I$ and $\delta H^I$. On eliminating
the $\delta \Psi^I$ and $\delta H^I$ terms, again the extra terms
can be rewritten as terms linear in $\delta Z^I, \delta g, \delta \phi^a, \bcn \delta Z^I$, which do not affect the ellipticity of the resulting
system.

\subsubsection{Topological Terms in $D=5$}

In $D=5$, we shall consider the topological term 
\be
S_{top} = \int  C_{IJK} A^I \w F^J \w F^K \ ,
\ee
where $C_{IJK}$ are constants. The gauge field equation, in the presence of the topological term, becomes 
\be
\nabla^{\mu} ( Q_{IJ}(\p) F^J_{\mu\nu}) - \frac{1}{4} C_{IJK} \epsilon_{\nu\mu\rho\sigma\lambda} F^{J\, \mu\rho} F^{K\, \sigma\lambda} = 0 \ . 
\ee
We show here the relevant changes to the equations of motion. 
The $-$ component of the gauge field equation, entering into the $\delta \Psi^I$ moduli fixing condition,  is modified by:
\be 
- C_{IJK} \epsilon^{+-ijk} Z^J_i H^K_{jk} \ ,  
\ee
which provides a linear contribution in $\d Z^I$. The $i$ component of the gauge field equation,
which enters into the construction of the elliptic system for $\delta Z^I$, is modified by:
\be
C_{IJK} \epsilon^{+-ijk} \bigg( - \P^J H^K_{jk} + 2r ( W^J_j - \P^J h_j ) Z^K_k \bigg) \ , 
\ee
which when linearized includes terms terms linear in $\delta Z^I$, $\delta \Psi^I$. On eliminating the $\delta \Psi^I$ term, the extra terms
can be rewritten as terms linear in $\delta Z^I, \delta g, \delta \phi^a, \bcn \delta Z^I$, which do not affect the ellipticity of the resulting
system.

\newsection{D=11 Supergravity}

The bosonic field content of D=11 supergravity is the D=11 metric ${\hat{g}}$, and a 4-form $F$, $F=dC$. The analysis of the moduli of transverse deformations of extremal horizons in
$D=11$ supergravity proceeds in a rather similar fashion to that of the Einstein-Maxwell-Dilaton theory. We shall again assume that the Killing vector ${\partial \over \partial u}$ is a symmetry of the full solution, i.e.
\be
\cL_{{\partial \over \partial u}} F = 0 \ ,
\ee
We decompose $F$ in Gaussian Null co-ordinates as
\bea
\label{4form}
F = du \wedge dr \wedge \Psi +r du \wedge W + dr \wedge Z + X
\eea
where $\Psi$ is a $u$-independent 2-form, $W, Z$ are $u$-independent 3-forms, and $X$ is a $u$-independent 4-form on $\cS$, which are all assumed to be analytic in $r$.

The Bianchi identity $dF=0$ then decomposes as
\bea
\td \Psi -W -r \dot{W}=0, \qquad \td W =0, \qquad \td Z - {\dot{X}}=0, \qquad \td X=0 \ .
\eea
The gauge field equations are given by
\bea
\label{g4eq}
\nabla^\nu F_{\nu \lambda_1 \lambda_2 \lambda_3}={q \over (4!)^2} \epsilon_{\lambda_1 \lambda_2 \lambda_3}{}^{\mu_1 \mu_2 \mu_3 \mu_4 \mu_5 \mu_6 \mu_7 \mu_8} F_{\mu_1 \mu_2 \mu_3 \mu_4} F_{\mu_5 \mu_6 \mu_7 \mu_8}
\eea
where $q$ is a constant. Here we have included a topological term in the action proportional to
$q F \wedge F \wedge C$. On imposing supersymmetry, the value of $q$ is fixed by
requiring consistency of the gauge field equations with the integrability conditions
of the gravitino Killing spinor equations. However, here our analysis is purely
in the bosonic sector, so the value of $q$ is kept arbitrary.

Finally, we have the Einstein equations:
\bea
\label{d11ein}
R_{\mu \nu}={1 \over 12} F_{\mu \lambda_1 \lambda_2 \lambda_3} F_\nu{}^{\lambda_1 \lambda_2 \lambda_3}
-{1 \over 144} {\hat{g}}_{\mu \nu} F_{\lambda_1 \lambda_2 \lambda_3 \lambda_4} F^{\lambda_1 \lambda_2 \lambda_3 \lambda_4} \ .
\eea
The decomposition of (\ref{g4eq}) and (\ref{d11ein}) in terms of the horizon data is provided in Appendix \ref{reduced11D_sugra}. 

\subsection{Moduli space computation}
\label{modcomp_11dsugra}
In this section we shall prove the finiteness of the black hole moduli space in $D=11$ supergravity by using the same procedure adopted for the Einstein-Maxwell-Dilaton theory. 
We must again identify the moduli associated with the gauge fields, in this case the 4-form $F$.
We shall adopt the following expansions for $\Psi$, $W$ and $X$ via
\bea
\Psi &=& {\bcc{{\Psi}}} + r \delta \Psi + {\cal{O}}(r^2)
\nonumber \\
W &=&  {\bcc{{W}}} + r \delta W + {\cal{O}}(r^2)
\nonumber \\
X &=& {\bcc{{X}}}+ r \delta X + {\cal{O}}(r^2) \ .
\eea
For the $Z$ moduli, we note that as $Z$ appears in $F$ as $dr \wedge Z$, which scales linearly with $r$, we shall expand out $Z$ as
\bea
Z= \delta Z + {\cal{O}}(r)
\eea
This expansion is consistent with requiring that under the transformation
\be
\label{diffeo2}
(u, r, y^A) \longrightarrow (\epsilon^{-1}u, \epsilon r , y^A ) \ , \qquad\qquad
\epsilon \in \mathbb{R}_{>0} \ .
\ee
all moduli terms are linear in $\epsilon$, as well as being required for
consistency with the $--$ Einstein equation. The gauge theory moduli are therefore
$\delta \Psi, \delta W, \delta X, \delta Z$. 

As done previously, we shall use the Bianchi identities and field equations to fix some of the
metric and gauge field moduli.
The Bianchi identity provides the following conditions:
\be
\label{Bianchi1}
\tilde{d} \d \P = 2 \d W \ , \\
\label{Bianchi2}
\tilde{d}\d Z = \d X \ ,
\ee
which we use to fix $\d W$ and $\d X$, respectively, in terms of $\delta \Psi$ and
$\delta Z$. 
By using the $-k_1k_2$ component of the gauge field equations (\ref{g4eq}), we further fix $\d \P$ as
\be
\notag
\label{psifix2}
\d \P_{k_1k_2} &=& \hhn^{\ell} \d Z_{\ell k_1k_2} - \hh^{\ell} Z_{\ell k_1k_2} + \d g_{k_1}{}^{\ell} \hP_{\ell k_2} - \d g_{k_2}{}^{\ell} \hP_{\ell k_1} \\
&-& \frac{1}{2} \d g_{\ell}{}^{\ell} \P_{k_1k_2} - \frac{q}{72} \epsilon_{k_1k_2}{}^{\ell_1\ell_2\ell_3\ell_4\ell_5\ell_6\ell_7} \d Z_{\ell_1\ell_2\ell_3} \hX_{\ell_4\ell_5\ell_6\ell_7} \ . 
\ee
Also, by using the $-i$ and $+-$ components of the Einstein equations (\ref{-iein}) and (\ref{+-ein}), we fix $\d h$ and $\d \D$ respectively as follows
\be
\label{hfix2}
\notag
\d h_i &=& \frac{1}{2} \hhn_i \d g_k{}^k - \frac{1}{2} \hhn^j \d g_{ji} + \frac{1}{2} \d g_{ij} \hh^j - \frac{1}{4} \hh_i \d g_k{}^k \\ 
&+& \frac{1}{4} \hP_{\ell_1\ell_2}\d Z_i{}^{\ell_1\ell_2} + \frac{1}{12} \d Z_{\ell_1\ell_2\ell_3} \hX_i{}^{\ell_1\ell_2\ell_3} \ , 
\ee
and
\be
\label{deltfix2}
\notag
\d \D &=& \frac{1}{3} \hhn_i \d h^i + \frac{1}{12} \hh^i \hhn_i \d g_k{}^k - \hh^i \d h_i - \frac{1}{6} \hD \d g_k{}^k - \frac{1}{12} \hh_i \hh^i \d g_k{}^k  + \frac{1}{3} \d g_{ij} \hh^i\hh^j \\
\notag
&-& \frac{1}{6} \d g_{ij} \hhn^i \hh^j - \frac{1}{6} \hh^j \hhn^i \d g_{ij} + \frac{1}{9} \d \P_{\ell_1\ell_2} \hP^{\ell_1\ell_2}  - \frac{1}{9}\hP^{\ell_1}{}_m \hP^{\ell_2 m} \d g_{\ell_1\ell_2} \\
\notag
&-& \frac{1}{108} (\hW - \hh\w \hP )_{\ell_1\ell_2\ell_3} \d Z^{\ell_1\ell_2\ell_3} + \frac{1}{216} \d X_{\ell_1\ell_2\ell_3\ell_4} \hX^{\ell_1\ell_2\ell_3\ell_4} \\
&-& \frac{1}{108}\hX^m{}_{\ell_1\ell_2\ell_3} \hX^{n\ell_1\ell_2\ell_3} \d g_{mn} - \frac{1}{216} (\hh\w \d Z)_{\ell_1\ell_2\ell_3\ell_4} \hX^{\ell_1\ell_2\ell_3\ell_4} \ . 
\ee
We remark that the expressions for $\delta \Psi$, $\delta X$ and $\delta h$ are linear in
$\delta Z, \delta g, \bcn \delta Z, \bcn \delta g$, whereas the expressions for
$\delta W$ and $\delta \Delta$ involve some second order derivatives acting
on $\delta g$, $\delta Z$. The remaining (unfixed) moduli are $\delta g$ and $\delta Z$.

We shall now proceed to find elliptic systems of PDEs constraining $\delta Z$ and $\delta g$,
starting with $\delta Z$. By taking the divergence of the Bianchi identity (\ref{Bianchi2}), we obtain 
\be
\label{divBianchi2}
\hhn^i \hhn_i \d Z_{k_1k_2k_3} - 3\hhn_{[k_1} \hhn^{\ell} \d Z_{ k_2 k_3] \ell} - 3{\bcc R}_{\ell [k_1} \d Z^{\ell}{}_{k_2k_3]} - \hhn^{\ell} \d X_{\ell k_1k_2k_3} = 0 \ ,
\ee
and by using the $k_1 k_2 k_3$ component of the gauge field equation (\ref{g4eq4}), we express the divergence of $\d X$ as follows
\be
\notag
\hhn^{\ell} \d X_{\ell k_1k_2k_3} &=& - 3 \hhn_{[k_1} \hhn^{\ell} \d Z_{k_2k_3] \ell} + 3 \hhn_{[k_1} ( \hh^{\ell} \d Z_{k_2k_3] \ell}) + 6 \hhn_{[k_1} ( \d g_{k_2}{}^{\ell} {\bcc{\P}}_{k_3] \ell})   \\
\notag
&+& \frac{3}{2} \hhn_{[k_1} ( {\bcc{\P}}_{k_2k_3]} \d g_{\ell}{}^{\ell} ) + \frac{q}{72} \epsilon_{[k_1 k_2}{}^{\ell_1\ell_2\ell_3\ell_4\ell_5\ell_6\ell_7} \hhn_{k_3]} ( \d Z_{\ell_1\ell_2\ell_3} \hX_{\ell_4\ell_5\ell_6\ell_7} ) \\
\notag
&+& \frac{3}{2} (\hhn_m \d g^{\ell}{}_{[k_1} + \hhn_{[k_1} \d g_{|m|}{}^{\ell} - \hhn^{\ell}\d g_{m [k_1} ) \hX^m{}_{k_2k_3]\ell} + \d g_{ij} \hhn^i \hX^j{}_{k_1k_2k_3}  \\
\notag
&+&   \hhn^m \d g_{m\ell} \hX^{\ell}{}_{k_1k_2k_3} - \frac{1}{2} \hhn^{\ell} \d g_m{}^m \hX_{\ell k_1k_2k_3} + \hhn^{\ell}( \hh \w \d Z )_{\ell k_1k_2k_3}\\
\notag
&-& 2 \hD \d Z_{k_1k_2k_3} -\d g_{ij} \hh^i \hX^j{}_{k_1k_2k_3} + 2 \d h^{\ell} \hX_{\ell k_1k_2k_3} + \hh^{\ell} \d X_{\ell k_1k_2k_3}  \\
\notag
&+& 2 (\d h\w \hP )_{k_1k_2k_3} + 2 (\hh \w \d \P )_{k_1k_2k_3} + 3 \d g^{\ell}{}_{[k_1} ( \hW - \hh \w \hP )_{k_2k_3]\ell} \\
\notag
&+& \hh^{\ell} \d X_{\ell k_1k_2k_3} - 2\hh^{\ell} ( \hh\w\d Z)_{\ell k_1k_2k_3} - \frac{1}{2} \d g_{\ell}{}^{\ell} ( \hW - \hh\w \hP )_{k_1k_2k_3} \\
\notag
&+& \frac{1}{2} ( -2\hh^{\ell} \d g_{\ell}{}^q + \hh^q\d g_{\ell}{}^{\ell} ) \hX_{q k_1k_2k_3} + 3 \hh^q \d g_{[k_1}{}^{\ell} \hX_{k_2k_3]\ell q} \\
\notag
&-& \frac{q}{24} \epsilon_{k_1k_2k_3}{}^{\ell_1\ell_2\ell_3\ell_4\ell_5\ell_6} ( \d \P_{\ell_1\ell_2} \hX_{\ell_3\ell_4\ell_5\ell_6} + \hP_{\ell_1\ell_2} \d X_{\ell_3\ell_4\ell_5\ell_6} - \hP_{\ell_1\ell_2} ( \hh\w \d Z )_{\ell_3\ell_4\ell_5\ell_6} ) \\
\notag
&+& \frac{q}{12} \d g_{\ell_1m} \epsilon_{k_1k_2k_3}{}^{\ell_1\ell_2\ell_3\ell_4\ell_5\ell_6} \hP^m{}_{\ell_2} \hX_{\ell_3\ell_4\ell_5\ell_6}  
+ \frac{q}{6} \d g_{\ell_3m} \epsilon_{k_1k_2k_3}{}^{\ell_1\ell_2\ell_3\ell_4\ell_5\ell_6} \hP_{\ell_1\ell_2} \hX^m{}_{\ell_4\ell_5\ell_6} \\
&+& \frac{q}{18} \epsilon_{k_1k_2k_3}{}^{\ell_1\ell_2\ell_3\ell_4\ell_5\ell_6} ( \hW - \hh\w \hP )_{\ell_1\ell_2\ell_3} \d Z_{\ell_4\ell_5\ell_6} \ .
\ee
On substituting this expression back into ({\ref{divBianchi2}}), 
the $\hhn_{[k_1} \hhn^{\ell} \d Z_{k_2k_3] \ell} $ terms cancel out. Furthermore, there
are a number of terms which are linear in $\delta X$, $\delta h$, $\delta \Psi$
which are eliminated on using the Bianchi identity ({\ref{Bianchi2}}), together with
({\ref{hfix2}}) and ({\ref{psifix2}}), producing terms which are linear in $\delta Z, \delta g, \bcn \delta Z, \bcn \delta g$. The resulting PDEs are an elliptic system for 
$\delta Z$, with principal symbol generated by $\bcn^2$.

Next, we consider the metric moduli $\delta g$. The linearized $ij$ component of the Einstein equations is
\be
\label{11dlin_ij}
\hhn^2 \d g_{ij} - \hhn_i \hhn_j \d g_k{}^k - ( \hhn_{\ell} \hhn_j - \hhn_j \hhn_{\ell} ) \d g^{\ell}{}_i - ( \hhn_{\ell} \hhn_i - \hhn_i \hhn_{\ell} ) \d g^{\ell}{}_j = \mathcal{C}_{ij} \ , 
\ee
where
\be
\notag
\mathcal{C}_{ij} &=& ( \hhn_i \d g^k{}_j + \hhn_j \d g^k{}_i - \hhn^k \d g_{ij} ) \hh_k -  \hhn_{(i} ( \hh_{j)} \d g_k{}^k ) + 2\hhn_{(i} (\d g_{j)k} \hh^k)  \\
\notag
&+& 2\hhn_{(i} (\hhn_{j)} \hp^b \hf_{ab} \d \p^a )  + 2\hhn_{(i} (\d Z^J_{j)}  \hQ_{IJ} \hP^I ) + 2 \hhn_{(i} ( \htH^J_{j)k}\hQ_{IJ} \d Z^{I\, k}  ) \\
\notag
&-&  8\hh_{(i} \d h_{j)} + 2( - \hD + \frac{1}{2} \hhn_k \hh^k - \hh_k \hh^k ) \d g_{ij} \\
\notag
&-& 2 \d g_{(i}{}^k \hhn_{|k|} \hh_{j)} - 2 \hh^k \hhn_{(i} \d g_{j)k} + 2 \hh^k \hhn_k \d g_{ij} - 2 \hh_{(i} \hhn^k \d g_{j)k} \\
\notag
&+& 4 \hh_k \hh_{(i} \d g_{j)}{}^k + 2 \hh_{(i} \hhn_{j)} \d g_k{}^k + \d g_k{}^k ( \hhn_{(i}\hh_{j)} - \hh_i \hh_j ) - 2 \d \P_{\ell (i} \hP_{j)}{}^{\ell} \\
\notag
&-& \hP_i{}^m \hP_j{}^n \d g_{mn} + (\hh\w \hP - \hW)_{\ell_1\ell_2 (i} \d Z_{j)}{}^{\ell_1\ell_2} + \frac{1}{3} \d X_{\ell_1\ell_2\ell_3 (i} \hX_{j)}{}^{\ell_1\ell_2\ell_3}  \\
\notag
&+& \frac{1}{2} \hX_{i\ell_1\ell_2}{}^m \hX_j{}^{\ell_1\ell_2 n} \d g_{mn}  - \frac{1}{3} (\hh\w \d Z)_{\ell_1\ell_2\ell_3 (i} \hX_{j)}{}^{\ell_1\ell_2\ell_3} - \frac{1}{3} \hg_{ij} \d \P_{\ell_1\ell_2} \hP^{\ell_1\ell_2} -  {1 \over 6} \delta g_{ij} \hP_{mn} \hP^{mn}
\\ \notag
&+&  {1 \over 3} \hg_{ij} \hP_m{}^\ell \hP^{mq} \delta g_{\ell q}
+{1 \over 72} \d g_{ij} \hX_{\ell_1 \ell_2 \ell_3 \ell_4} \hX^{\ell_1 \ell_2 \ell_3 \ell_4}
+ \frac{1}{36} \hg_{ij} \d X_{\ell_1\ell_2\ell_3\ell_4} \hX^{\ell_1\ell_2\ell_3\ell_4}
\\ \notag
 &-& \frac{1}{18} \hg_{ij} \hX^m{}_{\ell_1\ell_2\ell_3} \hX^{n \ell_1\ell_2\ell_3} \d g_{mn} 
+ \frac{1}{9} \hg_{ij} (\hW - \hh\w \hP)_{\ell_1\ell_2\ell_3} \d  Z^{\ell_1\ell_2\ell_3}
\\
 &-& \frac{1}{36} \hg_{ij} (\hh \w \d Z)_{\ell_1\ell_2\ell_3\ell_4} \hX^{\ell_1\ell_2\ell_3\ell_4} \ , 
\ee
which is linear in $\d Z, \d g, \hhn \d Z, \hhn \d g$. 
Furthermore, in ({\ref{11dlin_ij}}), terms of the form $( \hhn_{\ell} \hhn_j - \hhn_j \hhn_{\ell} ) \d g^{\ell}{}_i$ can be rewritten as terms linear in $\delta g$ and the Riemann tensor ${\bcc R}$, hence can be incorporated into the algebraic term on the RHS. The trace term $\delta g_k{}^k$
is again fixed by the elliptic condition ({\ref{trelliptic}}), so ({\ref{11dlin_ij}})
is an elliptic set of PDEs for the traceless part of $\delta g$, with principal symbol generated by $\bcn^2$.

Taken together, the conditions ({\ref{divBianchi2}}) and ({\ref{11dlin_ij}}) and ({\ref{trelliptic}}) constitute elliptic PDEs on the moduli $\{\delta Z, \delta g \}$,
with the remaining moduli $\{\delta W, \delta \Psi, \delta X, \delta h, \delta \Delta \}$
fixed in terms of $\{\delta Z,  \delta g \}$
by ({\ref{Bianchi1}}), ({\ref{Bianchi2}}), ({\ref{psifix2}}), ({\ref{hfix2}}) and ({\ref{deltfix2}}). The moduli space is therefore finite dimensional.

\section{Conclusions}

We have proven that the moduli space of transverse deformations 
of extremal event horizons in
Einstein-Maxwell-Dilaton theory in any dimension, and including
topological terms in four and five dimensions, is finite dimensional.
We have also demonstrated the same result in eleven-dimensional
supergravity, with an arbitrary coupling of the topological term.
The treatment of the gauge field moduli in both these theories is
very similar. We remark that some of the Einstein-Maxwell-Dilaton theories could
be obtained by dimensional reduction of pure gravity in higher dimensions.
In such cases, the finite dimensionality of the moduli spaces would be inherited
from the higher dimensional calculation via the result of \cite{Li:2015wsa}.
However, the case of D=11 supergravity cannot be obtained from such a reduction.

	There are a number of further theories for which it would be interesting to
investigate the moduli space of the transverse deformations, such as gravity coupled to
a non-abelian gauge theory. Horizons in such theories have been considered in
\cite{Li:2013gca}. Higher derivative theories, for example $\alpha'$-corrected heterotic
theory, could also be considered. Supersymmetric event horizons in this theory
were analyzed in \cite{Fontanella:2016aok}, and it is unclear how the higher derivative
terms would alter the elliptic systems. Beyond proving that the moduli space of
deformations is finite-dimensional, the next step is to actually count the moduli,
or at least to obtain further conditions on the number of moduli. 
Work in this direction is in progress.

\setcounter{section}{0}
\setcounter{subsection}{0}

\appendix{The field equations}
In this appendix we list the various components of the field equations,
expressed in terms of the horizon data. We denote by ${\tilde{\nabla}}$ the 
Levi-Civita connection on ${\cal{S}}$, restricted to $r=const.$.

\subsection{Einstein-Maxwell-Dilaton theory}
\label{reducedEMD}

The components of the Einstein equation (\ref{einst}) are:

{\flushleft{The $++$ component:}}
\be
\notag
&&R_{++}   - \frac{1}{2} r^2 Q_{IJ} ( W_i^I W^{J\, i} - 2W_i^{I} \P^{J} h^i + \P^I\P^J h_i h^i ) \\ 
&&- \frac{1}{4} r^3\D  Q_{IJ} ( 2W^{I}_i Z^{J\, i} - 2 \P^{I} Z^{J}_i h^i ) 
- \frac{1}{8} r^4 \D^2 (  f_{ab} \dot{\p}^a \dot{\p}^b + Q_{IJ} Z_i^I Z^{J\, i} ) = 0 \ .
\ee
The $+-$ component: 
\be
\label{EMDeins+-}
\notag
&&R_{+-} + \frac{1}{2(D-2)} Q_{IJ} ( 2 H^I_{ij} H^{J\, ij} + (D-3) \P^I\P^J) - \frac{1}{D-2} V\\
\notag
&&+r \frac{4-D}{2(D-2)} Q_{IJ} ( W_i^I Z^{J\, i} - \P^I Z^J_i h^i) \\
&&+r^2 \bigg( \frac{4-D}{4(D-2)} \D^2 Q_{IJ} Z^I_i Z^{J\, i} - \frac{1}{4}  \D f_{ab} \dot{\p}^a \dot{\p}^b \bigg) = 0 \ .
\ee
The $--$ component:
\be
R_{--} - \frac{1}{2} f_{ab} \dot{\p}^a \dot{\p}^b - \frac{1}{2} Q_{IJ} Z^I_i Z^{J\, i} = 0 \ .
\ee
The $+i$ component:
\be
\notag
&&R_{+i} + \frac{1}{2}r Q_{IJ} (\P^I W^J_i - \P^I\P^J h_i - H^I_{ij} W^{J\, j} + H^I_{ij} \P^J h^j )  \\
&& - \frac{1}{4} r^2 \D ( f_{ab} \dot{\p}^a \hn_i \p^b + Q_{IJ} H^I_{ij} Z^{J\, j} - Q_{IJ} \P^I Z^J_i ) + \frac{1}{4} r^3 \D h_i f_{ab} \dot{\p}^a \dot{\p}^b = 0  \ .
\ee
The $-i$ component:
\be
\label{EMDeins-i}
R_{-i} - \frac{1}{2} f_{ab} \dot{\p}^a \hn_i \p^b - \frac{1}{2} Q_{IJ} ( \P^I Z^J_i + Z^{I\, j}H^J_{ij} ) + \frac{1}{2} r h_i f_{ab} \dot{\p}^a \dot{\p}^b = 0 \ .
\ee
The $ij$ component:
\be
\label{ijeins}
\notag
&&R_{ij} - \frac{1}{2} f_{ab} \hn_i\p^a \hn_j \p^b - \frac{1}{2} Q_{IJ} H^I_{ik} H^J_j{}^k - \frac{1}{D-2} g_{ij} V \\
\notag
&&+ \frac{1}{4(D-2)} g_{ij} Q_{IJ} (H^I_{k\ell} H^{J\, k\ell} - 2 \P^I \P^J )\\
\notag
&&+ r \bigg( f_{ab} h_{(i} \dot{\p}^a \hn_{j)} \p^b + \frac{1}{D-2} g_{ij} Q_{IJ} Z^I_{\ell} ( W^{J\, \ell} - \P^J h^{\ell} ) -  Q_{IJ} ( W^I_{(i} Z^J_{j)} - \P^I Z^J_{(j} h_{i)}  )  \bigg) \\
&&+ r^2 \bigg( - \frac{1}{2} h_i h_j f_{ab} \dot{\p}^a \dot{\p}^b - \frac{1}{2} \D Q_{IJ} Z^I_i Z^J_j  + \frac{1}{2(D-2)} \D g_{ij} Q_{IJ} Z^I_{\ell} Z^{J\, \ell} \bigg) = 0 \ .
\ee
The components of the gauge field equation (\ref{gauge}) are:

{\flushleft{The $+$ component:}}
\be 
\label{gauge_1}
\notag&& - \hn^iW_i^I +  h^i W^I_i + \P^I \hn^i h_i - h^i h_i\P^I  - \frac{1}{2}\tilde{d}h_{ij} H^{I\, ij} \\
\notag
&& Q^{IK}\partial_a Q_{KJ}( \P^J h_i \hn^i \p^a -  W_i^J\hn^i \p^a) + h_i\hn^i\P^I \\
\notag 
&&+ r \bigg( Q^{IK}\partial_a Q_{KJ}( \dot{\p}^a h^i W_i^J -  h^2\P^J \dot{\p}^a 
  -\frac{1}{2}  \D Z^J_i \hn^i\p^a  - \frac{1}{2}  \D \dot{\p}^a \P^J ) \\
\notag
&& - \frac{1}{2} \D \dot{\P}^I 
 + h^i \dot{W}^I_i  + \frac{3}{2} \D h^i Z^I_i -  Z^I_i\hn^i \D - \frac{1}{2}\D\hn^iZ_i^I  - h^2\dot{\P}^I  - \P^I h^i \dot{h}_i\\
 \notag
&&  + ( \frac{1}{2} \dot{g}_k{}^k h_j - h^i \dot{g}_{ij} )  (W^{I\, j} - \P^I h^j ) - \frac{1}{4}  \dot{g}_k{}^k \D\P^I +  h_i \dot{h}_j H^{I\, ij} \bigg)\\
\notag 
&& + \frac{r^2}{2} \bigg( Q^{IK}\partial_a Q_{KJ} \dot{\p}^a \D  h^iZ^J_i + ( \frac{1}{2} \dot{g}_k{}^k h_j - h^i \dot{g}_{ij} ) \D Z^{I\, j} \\
&&+ 2 \dot{\D}h^iZ^I_i  + \D h^i\dot{Z}^I_i - \D \dot{h}^i Z^I_i   \bigg) = 0 \ .
\ee
The $-$ component: 
\be
\label{gauge_2}
\notag
&&  \dot{\P}^I  - \hn^i Z^I_i  + h^i Z_i^I - Q^{IK}\partial_a Q_{KJ}\hn^i \p^a Z^J_i + \frac{1}{2} \dot{g}_k{}^k \P^I + Q^{IK}\partial_a Q_{KJ}\dot{\p}^a \P^J  \\
&&+ r \bigg( Q^{IK}\partial_a Q_{KJ}  \dot{\p}^a h^i Z^J_i + h^i \dot{Z}^I_i 
+  \dot{h}^i Z^I_i + ( \frac{1}{2} \dot{g}_k{}^k h_j - h^i \dot{g}_{ij} ) Z^{I\, j}  \bigg) = 0 \ .
\ee
The $i$ component:
\be
\label{gauge_3}
\notag
&& W^I_i + \hn^j H^I_{ji} - \P^I h_i     - h^j H^I_{ji} + Q^{IK}\partial_a Q_{KJ}\hn^j \p^a H^J_{ji} \\
\notag
&& +r \bigg( Q^{IK}\partial_a Q_{KJ} ( \dot{\p}^a W^J_i -  \P^J\dot{\p}^a h_i -  \dot{\p}^a h^j H^J_{ji} ) + 2\D Z^I_i \\
\notag
&&+  \dot{W}^I_i -  \dot{\P}^I h_i  -  \P^I \dot{h}_i -  h^j \dot{H}^I_{ji}  -  \dot{h}^j H^I_{ji}  + \frac{1}{2}  \dot{g}_k{}^k ( W^I_i - \P^I h_i ) \\
\notag
&&  - \frac{1}{2}  \dot{g}_k{}^k h^j H^I_{ji}+  h^j \dot{g}_{jk} H^{I\, k}{}_i+  h_j H^{I\, jk}\dot{g}_{ki} -  \dot{g}_{ij} ( W^{I\, j} - \P^I h^j )  \bigg) \\
&&+ r^2 \bigg( Q^{IK}\partial_a Q_{KJ}\D  \dot{\p}^a Z^J_i   + \dot{\D}Z^I_i + \D \dot{Z}^I_i  + \frac{1}{2} \dot{g}_k{}^k \D Z^I_i   -  \dot{g}_{ij}  \D Z^{I\, j}   \bigg) = 0 \ .
\ee
The scalar field equation (\ref{scalar}) reduces in terms of the horizon data  to:
\be
\label{hscalar}
\notag
&&\hn_i \hn^i \p^a - h^i \hn_i \p^a + f^{ab}(\partial_c f_{bd} - \frac{1}{2}  \partial_b f_{cd} ) \hn_i \p^c \hn^i \p^d  \\
\notag
&&+\frac{1}{4} f^{ab} \partial_b Q_{IJ}  ( 2 \P^I \P^J  - H^I_{ij} H^{J\, ij} ) - f^{ab} \partial_b V\\
\notag
&& +r \bigg( \hn^j \p^a ( h^i \dot{g}_{ij} - \frac{1}{2} \dot{g}_k{}^k h_j )  + \dot{\p}^a (2\D + h_i h^i)  - \dot{\p}^a \hn^i h_i - 2 h_i \hn^i \dot{\p}^a \\
\notag
&& - f^{ab}(\partial_c f_{bd} + \partial_d f_{bc} - \partial_b f_{cd} ) \dot{\p}^c \hn_i \p^d  h^i -\frac{1}{2} f^{ab} \partial_b Q_{IJ} ( Z^I_i W^{J\, i} -  h^i Z^I_i \P^J ) \bigg) \\
\notag
&&+ r^2 \bigg( -  h^j \dot{\p}^a( h^i \dot{g}_{ij} - \frac{1}{2} \dot{g}_k{}^k h_j ) + (\D + h^i h_i) \ddot{\p}^a  + \dot{\D} \dot{\p}^a + 2 h_i \dot{h}^i \dot{\p}^a + \frac{3}{4} \dot{g}_k{}^k \D \dot{\p}^a \\
&&+ f^{ab}(\partial_c f_{bd} - \frac{1}{2}  \partial_b f_{cd} ) \dot{\p}^c \dot{\p}^d (\D + h^i h_i)  -\frac{1}{4} f^{ab} \partial_b Q_{IJ}\D Z^I_i Z^{J\, i} \bigg)  = 0 \ .
\ee

\subsection{D=11 Supergravity}
\label{reduced11D_sugra}
The gauge field equation ({\ref{g4eq}}) decomposes into the following components. 

{\flushleft{The $+-k$ component:}}
\bea
\label{g4eq1}
\tn^\ell \Psi_{\ell k} -r h^\ell {\dot{\Psi}}_{\ell k}
-{1 \over 2}r (\td h)^{mn} Z_{mnk} +r h^\ell {\dot g}_\ell{}^q \Psi_{qk}
-{1 \over 2}r h^q \Psi_{qk} {\dot{g}}_m{}^m +r h^\ell {\dot g}_k{}^q \Psi_{\ell q}
\nonumber \\
= {q \over 576} \epsilon_k{}^{\ell_1 \ell_2 \ell_3 \ell_4 \ell_5 \ell_6 \ell_7 \ell_8} X_{\ell_1 \ell_2 \ell_3 \ell_4}
X_{\ell_5 \ell_6 \ell_7 \ell_8}-{qr \over 72}
\epsilon_k{}^{\ell_1 \ell_2 \ell_3 \ell_4 \ell_5 \ell_6 \ell_7 \ell_8} h_{\ell_1} Z_{\ell_2 \ell_3 \ell_4} X_{\ell_5 \ell_6 \ell_7 \ell_8} \ .
\eea

{\flushleft{The $+k_1 k_2$ component:}}
\bea
\label{g4eq2}
\tn^\ell (W-h \wedge \Psi)_{\ell k_1 k_2}+
2 r \Delta h^\ell Z_{\ell k_1 k_2}
+h^\ell (W-h \wedge \Psi)_{\ell k_1 k_2}
\nonumber \\
+r h^\ell({\dot{W}}-{\dot{h}}\wedge \Psi - h \wedge {\dot{\Psi}})_{\ell k_1 k_2}
+{1 \over 2}(-\td h^{mn}+r h^m {\dot{h}}^n) X_{mn k_1 k_2}
\nonumber \\
+{1 \over 2} r\td h^{mn}(h \wedge Z)_{mn k_1 k_2}
+{1 \over 2} r (-2 h^\ell {\dot{g}}_\ell{}^q +h^q {\dot{g}}_m{}^m ) (W - h \wedge \Psi)_{q k_1 k_2}
\nonumber \\
-r h^\ell {\dot{g}}_{k_1}{}^q (W-h \wedge \Psi)_{\ell q k_2}
+r h^\ell {\dot{g}}_{k_2}{}^q (W-h \wedge \Psi)_{\ell q k_1}
\nonumber \\
={q \over 72} \epsilon_{k_1 k_2}{}^{\ell_1 \ell_2 \ell_3 \ell_4 \ell_5 \ell_6 \ell_7} (W-h \wedge \Psi)_{\ell_1 \ell_2 \ell_3}
X_{\ell_4 \ell_5 \ell_6 \ell_7}
-{qr \over 18} \epsilon_{k_1 k_2}{}^{\ell_1 \ell_2 \ell_3 \ell_4 \ell_5 \ell_6 \ell_7} h_{\ell_1} Z_{\ell_2 \ell_3 \ell_4} W_{\ell_5 \ell_6 \ell_7} \ .
\nonumber \\
\eea

{\flushleft{The $-k_1 k_2$ component:}}
\bea
\label{g4eq3}
\tn^\ell Z_{\ell k_1 k_2} &=& {\dot{\Psi}}_{k_1 k_2}
+h^\ell Z_{\ell k_1 k_2}
-{\dot{g}}_{k_1}{}^\ell \Psi_{\ell k_2}
+{\dot{g}}_{k_2}{}^\ell \Psi_{\ell k_1}
+{1 \over 2}{\dot{g}}_m{}^m \Psi_{k_1 k_2}
\nonumber \\
&+&{q \over 72} \epsilon_{k_1 k_2}{}^{\ell_1 \ell_2 \ell_3 \ell_4 \ell_5 \ell_6 \ell_7} Z_{\ell_1 \ell_2 \ell_3}
X_{\ell_4 \ell_5 \ell_6 \ell_7}
\eea
and the $k_1 k_2 k_3$ component:
\bea
\label{g4eq4}
\tn^\ell (X-r h \wedge Z)_{\ell k_1 k_2 k_3}
+2r \Delta Z_{k_1 k_2 k_3}
-(h^\ell+r {\dot{h}}^\ell) X_{\ell k_1 k_2 k_3}
+(W-h \wedge \Psi)_{k_1 k_2 k_3}
\nonumber \\
+r (\dot{W}-{\dot{h}} \wedge \Psi -h \wedge {\dot \Psi})_{k_1 k_2 k_3}
-3r {\dot{g}}^\ell{}_{[k_1} (W-h \wedge \Psi)_{k_2 k_3] \ell}
-r h^\ell {\dot{X}}_{\ell k_1 k_2 k_3}
\nonumber \\
+2r h^\ell(h \wedge Z)_{\ell k_1 k_2 k_3}
+{1 \over 2} r {\dot{g}}_m{}^m (W-h \wedge \Psi)_{k_1 k_2 k_3}
\nonumber \\
-{1 \over 2} r(-2 h^\ell {\dot{g}}_\ell{}^q + h^q {\dot{g}}_m{}^m)
X_{q k_1 k_2 k_3} -3r h^q {\dot{g}}_{[k_1}{}^\ell X_{k_2 k_3] \ell q}
\nonumber \\
=-{q \over 24} \epsilon_{k_1 k_2 k_3}{}^{\ell_1 \ell_2 \ell_3 \ell_4 \ell_5 \ell_6} \Psi_{\ell_1 \ell_2} (X-r h \wedge Z)_{\ell_3 \ell_4 \ell_5 \ell_6}
+{qr \over 18} \epsilon_{k_1 k_2 k_3}{}^{\ell_1 \ell_2 \ell_3 \ell_4 \ell_5 \ell_6} (W-h \wedge \Psi)_{\ell_1 \ell_2 \ell_3}
Z_{\ell_4 \ell_5 \ell_6} \ .
\nonumber \\
\eea

It should be noted that in ({\ref{g4eq1}})-({\ref{g4eq4}}), we have supppressed
the appearance of terms of the form $\dot{Z}$, and also
$\dot{\Delta} Z, {\dot{h}} Z, {\dot{g}} Z$, because as we explain in section \ref{modcomp_11dsugra}, $Z$ is linear in the moduli, and hence these
terms are suppressed in the moduli space calculation.
Furthermore, ({\ref{g4eq2}}) has been simplified by making
use of ({\ref{g4eq3}}) to eliminate the $\tn^\ell Z_{\ell k_1 k_2}$ term from ({\ref{g4eq2}}).

The Einstein equation (\ref{d11ein}) decomposes into the following components:

{\flushleft{The $++$ component:}}
\be
R_{++} = \frac{1}{12} r^2 (W- h\w \P)_{\ell_1\ell_2\ell_3} (W - h\w \P + r \D Z)^{\ell_1\ell_2\ell_3}   \ .
\ee
The $+-$ component:
\be
\label{+-ein}
\notag
R_{+-} &=& - \frac{1}{6} \P_{\ell_1\ell_2} \P^{\ell_1\ell_2} + \frac{1}{36} r (W - h\w \P )_{\ell_1\ell_2\ell_3} Z^{\ell_1\ell_2\ell_3} \\
&-& \frac{1}{144}  X_{\ell_1\ell_2\ell_3\ell_4} X^{\ell_1\ell_2\ell_3\ell_4} + \frac{1}{72} r (h\w Z)_{\ell_1\ell_2\ell_3\ell_4} X^{\ell_1\ell_2\ell_3\ell_4} \ .
\ee
The $+i$ component:
\be
\notag
R_{+i} &=& - \frac{1}{4} r \P_{\ell_1\ell_2} (W- h\w \P + \frac{1}{2} r \D Z)_i{}^{\ell_1\ell_2} - \frac{1}{12} r^2 (W - h\w \P)_{\ell_1\ell_2\ell_3} (h\w Z)_i{}^{\ell_1\ell_2\ell_3} \\
&+& \frac{1}{12} r X_i{}^{\ell_1\ell_2\ell_3} ( W -h\w \P + \frac{1}{2} r \D Z )_{\ell_1\ell_2\ell_3} \ .
\ee
The $-i$ component:
\be
\label{-iein}
R_{-i} = \frac{1}{4} \P_{\ell_1\ell_2} Z_i{}^{\ell_1\ell_2} + \frac{1}{12} Z_{\ell_1\ell_2\ell_3} X_i{}^{\ell_1\ell_2\ell_3} \ .
\ee
The $ij$ component:
\be
\notag
R_{ij} &=& - \frac{1}{2} \P_{i\ell} \P_j{}^{\ell} + \frac{1}{2} r (W - h\w\P + \frac{1}{2} r \D Z)_{\ell_1\ell_2 (i} Z_{j)}{}^{\ell_1\ell_2} + \frac{1}{12} X_{i\ell_1\ell_2\ell_3} X_j{}^{\ell_1\ell_2\ell_3} \\
\notag
&+& \frac{1}{6} r (h\w Z)_{\ell_1\ell_2\ell_3 (i} X_{j)}{}^{\ell_1\ell_2\ell_3} + \frac{1}{12} g_{ij} \P_{\ell_1\ell_2} \P^{\ell_1\ell_2} - \frac{1}{144} g_{ij} X_{\ell_1\ell_2\ell_3\ell_4} X^{\ell_1\ell_2\ell_3\ell_4} \\
&-& \frac{1}{18} r g_{ij}(W - h\w \P )_{\ell_1\ell_2\ell_3} Z^{\ell_1\ell_2\ell_3}  + \frac{1}{72} r g_{ij} (h\w Z)_{\ell_1\ell_2\ell_3\ell_4} X^{\ell_1\ell_2\ell_3\ell_4} \ .
\ee

\appendix{Ricci Tensor}
The Ricci tensor components in the frame basis ({\ref{nhbasis}}) are 
\be
\notag
R_{++} &=& r^2 \bigg( \frac{1}{2}\hn_i \hn^i \D  -\frac{3}{2} h^i \hn_i \D - \frac{1}{2} \D \hn_i h^i + \D h^i h_i + \frac{1}{4} \tilde{d}h_{ij} \tilde{d}h^{ij} \bigg) \\
\notag
&+& r^3 \bigg( -  h^i \hn_i \dot{\D} - \frac{1}{2} \dot{\D} \hn_i h^i + \frac{1}{2} \dot{h}^i \hn_i \D + \frac{1}{2} \D \hn_i \dot{h}^i + 2 \dot{\D} h^i h_i - \D h^i \dot{h}_i \\
&+&  \frac{1}{4}\D \dot{g}_k{}^k h_i h^i - \frac{1}{2} \D \dot{g}_{ij} h^i h^j  - \frac{1}{4} \dot{g}_k{}^k h_i \hn^i \D  + \frac{1}{2} \dot{g}_{ij} h^i \hn^j\D- h^i \dot{h}^j \tilde{d}h_{ij} \bigg) + {\cal{O}}(2) \ ,
\nonumber \\
\ee
\be
\notag
R_{+-} &=& \frac{1}{2}\hn_i h^i  - \D - \frac{1}{2} h^ih_i \\ 
&+& r \bigg( \frac{1}{2}\hn_i \dot{h}^i - \frac{1}{2}\D \dot{g}_k{}^k  - \frac{1}{4} \dot{g}_k{}^k h_i h^i + \frac{1}{2} \dot{g}_{ij} h^i h^j -2\dot{\D} - 2 h^i \dot{h}_i  \bigg) + {\cal{O}}(2) \ ,
\ee
\be
R_{--} ={\cal{O}}(2)  \ ,
\ee
\be
\notag
R_{+i} &=&  r \bigg( \frac{1}{2}\hn^k \tilde{d}h_{ik} +  h^j \tilde{d}h_{ji} + \D h_i - \hn_i \D   \bigg) \\
\notag
&+& r^2 \bigg(-\frac{1}{2} \D \dot{h}_i  + \frac{1}{2} \dot{h}_i \hn_j h^j + h^j \hn_j \dot{h}_i  - \frac{1}{2} h_i \hn_j \dot{h}^j - \frac{1}{2} \hn_i (h_j \dot{h}^j )\\
\notag
&+& 2\dot{\D} h_i - \frac{1}{2} \hn_i \dot{\D} + 3 h^j h_{[i} \dot{h}_{j]} - \frac{3}{4} \D \dot{g}_{ij} h^j  +  \frac{1}{2} \dot{g}_{ij} \hn^j \D + \frac{1}{4} \D \hn^j\dot{g}_{ij}   \\
&-&  \frac{1}{4} \hn_i (\D  \dot{g}_k{}^k) + \frac{3}{8}\D h_i \dot{g}_k{}^k  + \frac{1}{2} \dot{g}_i{}^j \td h_{jk} h^k + \frac{1}{4} \dot{g}_k{}^k h^j \td h_{ji} + \frac{1}{2} \td h_i{}^j \dot{g}_{jk} h^k   \bigg) + {\cal{O}}(2)  \ , 
\nonumber \\
\ee
\be
R_{-i} &=& \dot{h}_i + \frac{1}{2} \hn^j \dot{g}_{ji} - \frac{1}{2} \hn_i \dot{g}_k{}^k + \frac{1}{4} h_i \dot{g}_k{}^k - \frac{1}{2} \dot{g}_{ij} h^j + {\cal{O}}(2) \ ,
\ee
\be
\notag
R_{ij} &=& {\tilde{{\mathcal R}}}_{ij}
+\tn_{(i} h_{j)} -{1 \over 2} h_i h_j
\nonumber \\
&+& r \bigg(\tn_{(i} \dh_{j)}-3h_{(i} \dh_{j)}
+\big(-\Delta+{1 \over 2}\tn_k h^k -h_k h^k\big)
{\dot{g}}_{ij}
-{\dot{g}}_{(i}{}^k \tn_{|k|} h_{j)}
\nonumber \\
&-& h^k \tn_{(i} {\dot{g}}_{j) k}
+h^k \tn_k {\dot{g}}_{ij}
-h_{(i} \tn^k {\dot{g}}_{j) k} +2 h_k h_{(i} {\dot{g}}_{j)}{}^k+ h_{(i} \tn_{j)} {\dot{g}}_k{}^k
\nonumber \\
&+&{1 \over 2}{\dot{g}}_k{}^k \big(\tn_{(i} h_{j)}-h_i h_j \big)   \bigg) + {\cal{O}}(2) \ .
\ee

Here ${\cal{O}}(2)$ consists of terms linear in ${\ddot{h}}, {\ddot{\Delta}}, {\ddot{g}}$,
and terms quadratic in ${\dot{h}}, {\dot{\Delta}}, {\dot{g}}$, which play no role in the moduli
space calculations.

\vskip 0.5cm
\noindent{\bf Acknowledgements} \vskip 0.1cm
\noindent  
 AF is partially supported by the EPSRC grant FP/M506655. JG is supported by the STFC Consolidated Grant ST/L000490/1. The authors would like to thank Carmen Li for useful discussions.
\vskip 0.5cm

\vskip 0.5cm
\noindent{\bf Data Management:} \vskip 0.1cm
\noindent  No data beyond those presented and cited in this work are needed to validate this study.

\end{document}